\documentclass[twocolumn,10pt]{asme2ej}

\usepackage{helvet} 
\title{\fontfamily{phv}\selectfont{\Huge{Title of the Article}}} 
\usepackage{epsfig} 

\title{Effect Of Running Shoes On Foot Impact During Running}

\author{Henry Nassif
    \affiliation{
	Department of Mechanical Engineering\\
	Massachusetts Institute of Technology\\
	Cambridge, MA 02139\\
    Email: hnassif@mit.edu
    }	
}

\author{Barbara Hughey 
    \affiliation{ Co-supervisor \\
    	Ph.D, Lab Manager and Instructor\\
        Department of Mechanical Engineering\\
    	Massachusetts Institute of Technology\\
		Cambridge, MA 02139\\
    	Email: bhughey@mit.edu
    }
}

\author{Peter So
    \affiliation{ Co-supervisor \\
    Professor of Mechanical Engineering and Biological Engineering \\ 
    Massachusetts Institute of Technology\\
	Cambridge, MA 02139\\
    Email: ptso@mit.edu
    }
}

\begin{document}

\maketitle    

\begin{abstract}
{\it Running is part of almost every sport, and requires a great amount of stamina, endurance, mental toughness and overall strength. At every step, the foot experiences ground reaction forces necessary to support the motion of the body. With the advancements in shoe technology, running shoes have grown in popularity among runners, as well as non-runners, because they reduce the risk of injuries from the impact felt by the foot. The purpose of this report is to analyze the effect of running shoes on impact forces on the foot. This is achieved through the use of three force pads fixed at different locations on the foot The force measured by each sensor is then used to estimate the vertical ground reaction force, using the sensors' calibrations equations . Based on the ground reaction force, the effective mass corresponding to the momentum change occurring during the transient phase of the impact is estimated. The results show that running at 9 miles per hour without running shoes generates an effective mass of \( M_{eff} = (14.9 \pm 4.8)\%M_{body} \) while running at the same speed with running shoes generates an effective mass of \( M_{eff} = (7.8 \pm 1.5 )\%M_{body} \) . The values highlight a significant reduction in the risk of pain or injuries accomplished by wearing running shoes.    
}
\end{abstract}

\begin{nomenclature}
\entry{$I$}{Impulse}
\entry{$F$}{force}
\entry{\( M_{body} \)}{Body Mass}
\entry{\( V_{com} \)}{Vertical Speed of the Center of Mass}
\entry{\( V_{foot} \)}{Vertical speed of the foot just before impact}
\entry{$g$}{Acceleration due to gravity at the Earth's surface}
\end{nomenclature}


\section{Introduction}

Running is a sport that has grown in popularity 
over the years due to its ease of access and the fact that it is relatively inexpensive. Most kinds of physical exercises are beneficial for remaining healthy, but many people prefer running over other forms of physical exercise. Beside physical exercise, the ability to run fast is crucial in many sports for passing a defender, catching a ball, or developing enough take off velocity for a jump.  Running is also one of the most intense physical activities, constantly involving impact between foot and ground. In the long run, the impact can cause injuries and pain. The impact force depends on the type of the collision, which can occur in 3 ways \cite{novacheck1998biomechanics} : a rear-foot strike (RFS), with the heel landing first; a mid-foot strike (MFS), with the heel and ball of the foot landing simultaneously; and a forefoot strike (FFS), with the ball of the foot landing before the heel. Approximately 80\% of contemporary shod distance runners are rear foot strikers whereas most professional sprinters are forefoot strikers \cite{lieberman2010foot}. Most of the remainder are characterized as midfoot strikers \cite{marti1989relationship}. 
 
 	With recent developments in shoe technology, it has become indispensable for any dedicated runner to possess a pair of running shoes, which provides a well-shaped fit tailored to one’s foot. The cushioning material in a running shoe protects the runner against the effects of repeated impacts \cite{cooper1986analysis}.Whether running shoes are necessary has received some heated debate over the past year, with studies emerging questioning the need for expensive running shoes 
 
The goal of this report is to compare the effect of running shoes on the impact forces experienced by the foot. Voltage is measured across calibrated force pads distributed along the foot to determine the shape and magnitude of the ground reaction forces. 
 
    The next section of this report analyses the basic theory behind the concepts and notions referred to in this experiment, such as gait cycle, ground reaction force and impulse. The following part outlines the apparatus as well as the experimental procedure followed to measure the force distribution on the foot. Finally, the results are presented and discussed as well as compared to historical data.

\section{The Mechanics Of Walking And Running}

The background for the experiment involves understanding the stance phase and gait cycle, the ground reaction force, the force impulse and force sensing resistors

\subsection{Stance Phase And Gait Cycle}

\begin{figure*}[!t]
  \centering
  \includegraphics[width=\linewidth]{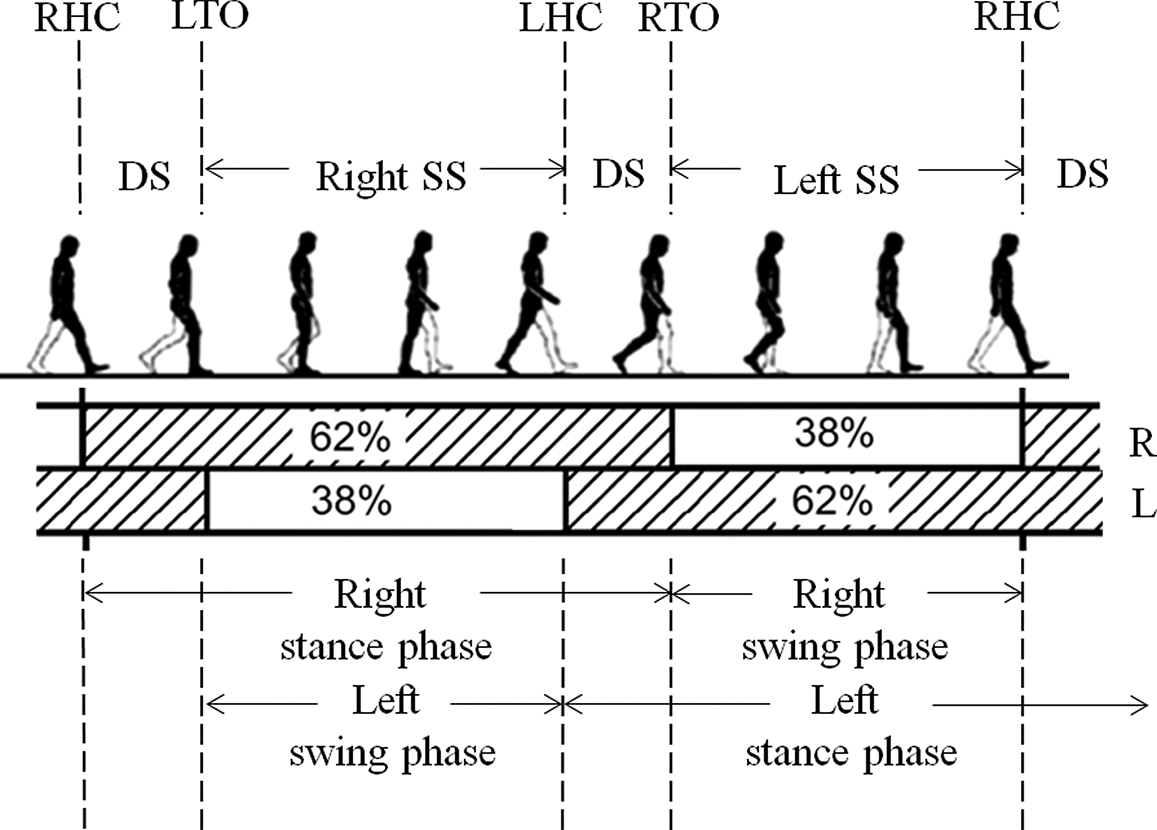}
  \caption{Phases of a single human gait cycle \cite{mummolo2013quantifying}.}
  \label{fig:gait}
\end{figure*}

The gait cycle is the periodic alternating motion of both legs governing the forward movement of the body. The cycle starts when one foot comes in contact with the ground and ends when the same foot contacts the ground again \cite{novacheck1998biomechanics}. The stance period of the gait cycle is the duration for which one foot is in contact with the ground \cite{novacheck1998biomechanics}. There are two stance phases per gait cycle, one for each leg. Toe off signals the start of the swing phase of the gait cycle, corresponding to the duration for which the foot is unsupported by the ground. Figure\ref{fig:gait} shows the different phases of the human gait cycle.

Because both feet are never off the ground at the same time during walking, the stance phases in walking last longer than half the gait cycle and always start with periods of double support, when both feet are on the ground. These periods occur at the beginning and at the end of each stance phase. The demarcation between walking and running occurs when periods of double support are replaced by periods of double float (airborne) at the beginning and the end of the swing phase. During running, there are no periods when both feet are in contact with the ground, since toe off occurs before the mid of the gait cycle. The timing of toe off depends on speed. Less time is spent in stance as the athlete moves faster. 

\subsubsection{Newton's Third Law And The Ground Reaction Force}

\begin{figure}[!t]
  \includegraphics[width=\linewidth]{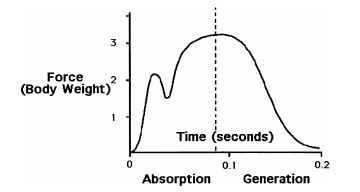}
  \caption{VGRF on foot during running at 3.5 m/s \cite{novacheck1998biomechanics}.}
  \label{fig:force}
\end{figure}

During the swing phase of the gait cycle no external forces are applied on the foot aside from wind resistance and gravity. On the contrary, during the stance phase, the ground constantly exerts a force on the foot. This force can be measured using force pad sensors and Newton's third law as follows: \textit{''When one body exerts a force on a second body, the second body simultaneously exerts a force equal in magnitude and opposite in direction to that of the first body''}. The downward force produced by stepping on the ground is thus matched by an upward force from the ground on the foot. Neglecting the anteroposterior reaction force in the direction of motion and the medial-lateral reaction force into the plane of motion, the vertical reaction force is the maximum ground reaction force denoted \textit{VGRF} \cite{tongen2010biomechanics}. Figure~\ref{fig:force} shows the VGRF for a single rear foot strike during running.  

The running VGRF of a RFS runner exhibits two peaks: one main peak preceded by a sizeable impact peak. The first (impact) peak corresponds to the force generated when landing on the heel and the second (main) peak corresponds to the force generated to propel the body upward. The dip in between both peaks is due to the center of gravity of the body changing the direction of its motion from downward during landing to upward during propulsion \cite{tongen2010biomechanics}.  
 
\begin{figure}[!t]
  \includegraphics[width=\linewidth]{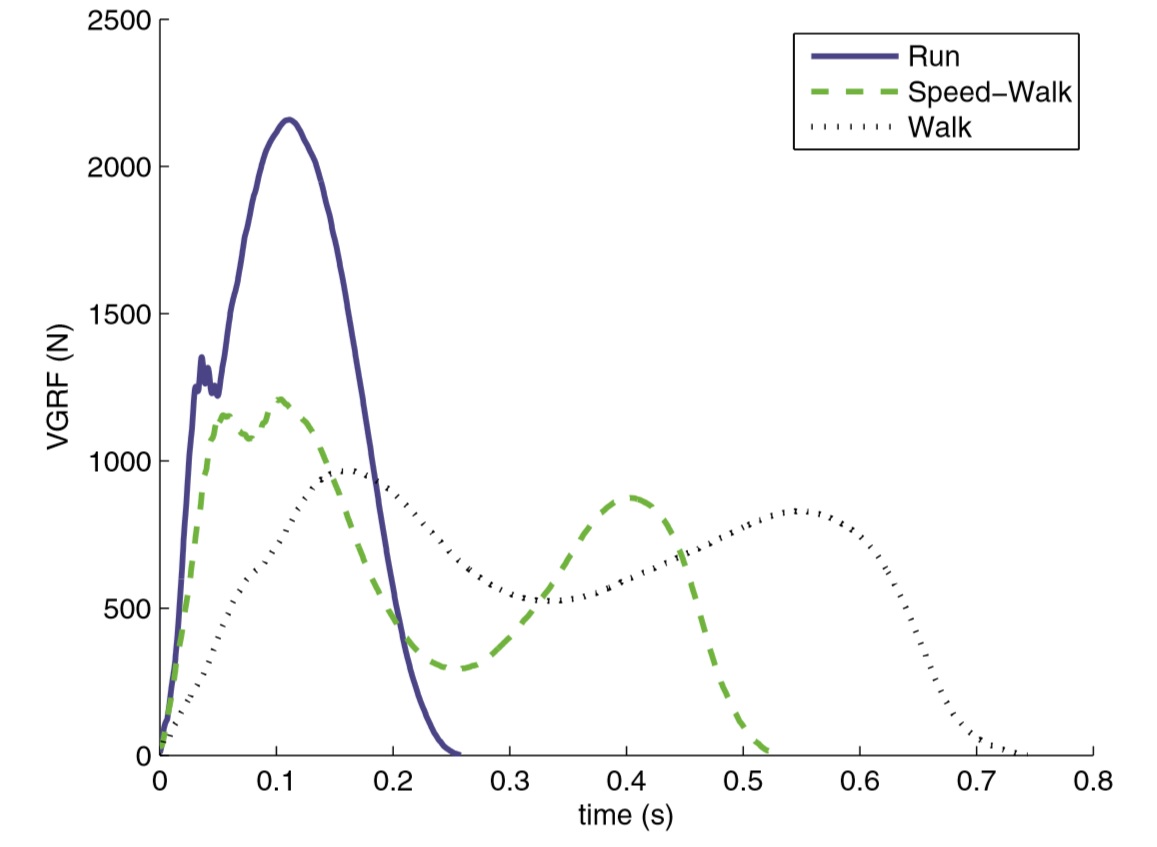}
  \caption{VGRF for one step during walking, speed-walking and running \cite{tongen2010biomechanics}.}
  \label{fig:force-compare}
\end{figure}

In order to run faster, the duration of the stance period of the gait cycle must be reduced \cite{tongen2010biomechanics}. This generates higher peak forces to support and propel the bodyweight in shorter amounts of time. As a result, running is characterized by high magnitude forces applied over short periods of contact, whereas walking is characterized by lower magnitude forces applied over longer periods of contact. This is illustrated by Fig.~\ref{fig:force-compare}.  

\subsubsection{Force Impulse}

During the stance period of the gait cycle, a runner repeatedly experiences the transient impact of the vertical ground reaction force, corresponding to the first peak observed in Fig.~\ref{fig:force}. This force ranges from 1.5 to as much as 3 times the subject’s body weight \cite{revill2008variability}.  The impulse of a step is the time integral of the force   over the duration  during which the force is applied, and is given by Equation~\ref{eq_one}.  

\begin{equation}
I = \int_{0}^{T} F dt 
\label{eq_one}
\end{equation}

The impulse is also equal to the change in the body's momentum during this period and is constant regardless of the method of running (RFS, MFS, and FFS) \cite{lieberman2010foot}.  While RFS runners are exposed to more injuries due to sudden forces with high magnitudes,  FFS runners minimize this risk by avoiding high force peaks through generating ground reaction forces without a distinct impact transient. 
  
The deceleration that happens during the transient part of the ground reaction force can be simulated by a fraction of the body mass slowing down until complete stop \cite{lieberman2010foot}. This mass is defined as the foot mass participating during impact following the initial foot-ground contact. At a specific speed, the lower the effective mass, the smaller is the risk of injuries \cite{novacheck1998biomechanics}. The relation between   and the impulse $I$ is given in Equation~\ref{eq_two} \cite{lieberman2010foot}, where $F(t)$ is the VGRF, $0-$ is the instant of time before impact, $T$ is the duration of the impact transient (time between the start of the transient phase and the peak transient force), \( M_{body} \) is the body mass, \( V_{com} \) is the vertical speed of the center of mass, \( V_{foot} \) is the vertical speed of the foot just before impact and $g$ is the acceleration due to gravity at the Earth's surface. 
 
\begin{equation}
I = \int_{0-}^{T} F dt = M_{body}(\Delta_{com}+gT) = M_{eff}(-V_{foot} + gT)
\label{eq_two}
\end{equation}
 
Given a force $F$, a foot vertical velocity \( V_{foot} \) and the duration of the impulse $T$, one can calculate the effective mass  associated with the collision using Equation~\ref{eq_three}.

\begin{equation}
M_{eff} = \frac{\int_{0-}^{T} F dt}{-V_{foot+gT}}
\label{eq_three}
\end{equation}

\subsubsection{Force Sensing Resistor}

\begin{figure*}[!t]
  \includegraphics[width=\linewidth]{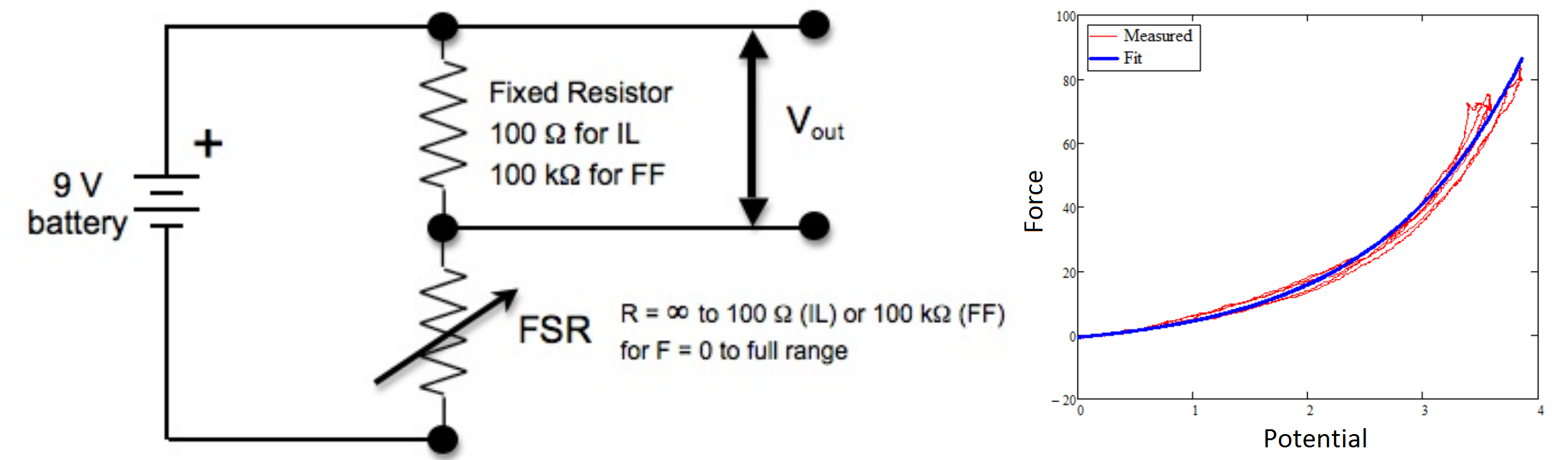}
  \caption{voltage divider circuit (left) and sample calibration curve for an IL-406 FSR (right).}
  \label{fig:voltage-divider}
\end{figure*}

In order to measure high forces, a voltage divider is assembled with a fixed resistor, the force pad, and a 9V battery. For voltages under 4 volts, the calibration curve is a natural exponential with large uncertainties \cite{exp2013}. This uncertainty in force increases at higher forces, since the force vs. voltage curve becomes steeper \cite{exp2013}. Figure~\ref{fig:voltage-divider} shows the voltage divider circuit as well as sample calibration curve.

\section{Measuring Force}

\begin{figure}[!t]
  \includegraphics[width=\linewidth]{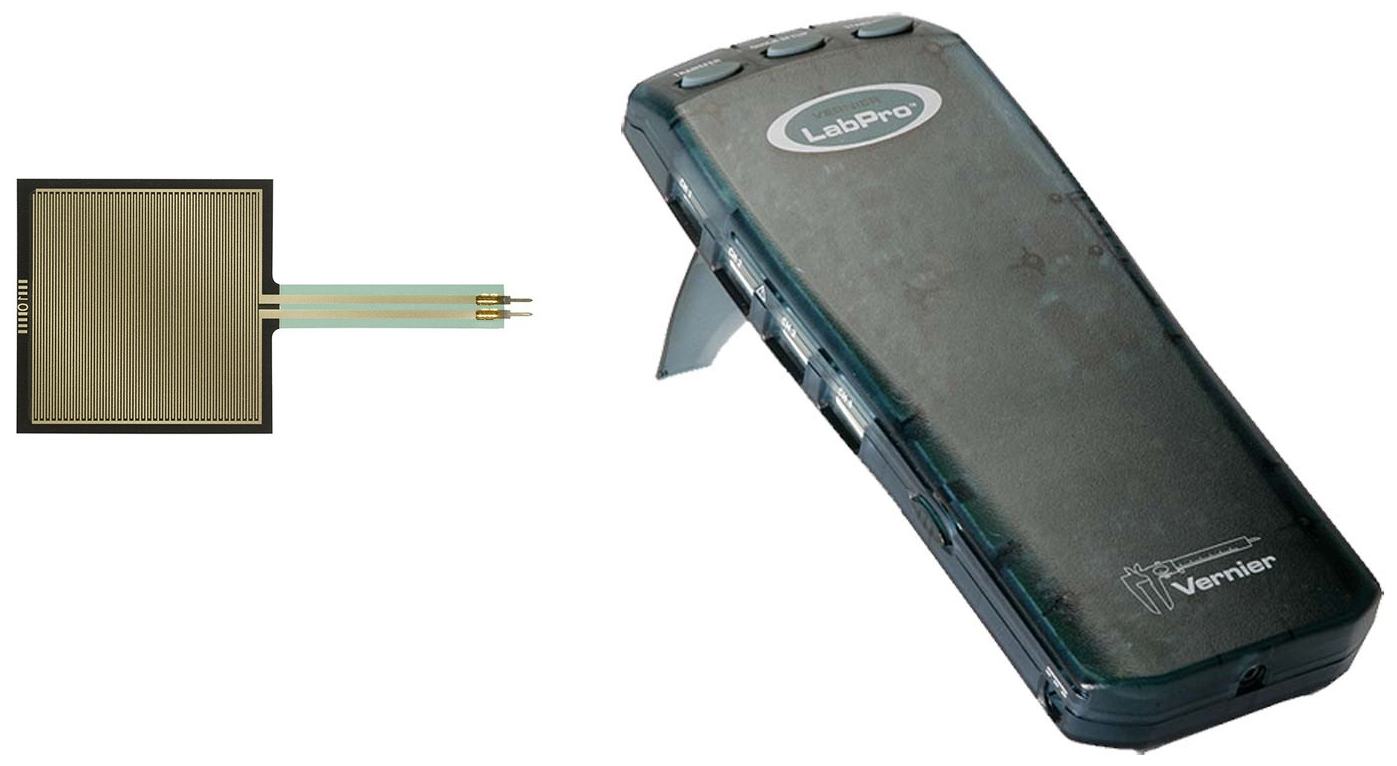}
  \caption{Force Pad Sensors (left) and Vernier LabPro DAQ (right)  .}
  \label{fig:force-pad-vernier}
\end{figure}

The apparatus for measuring the force distribution along the foot included three \textit{Interlink 406 Force Sensing resistors (IL-406)} and a \textit{Vernier LabPro} Data acquisition interface. The sampling rate was set to 400 samples/second. Figure~\ref{fig:force-pad-vernier} shows the instruments used.  

\subsection{Calibrating The Force Pads}

The calibration was done against a Vernier force plate by plotting the force with respect to the sensor voltage and fitting a natural exponential to the data. Equation~\ref{eq_four} shows the generic form of the expression for the force measured by each sensor as a function of the sensor voltage, where $A$,$B$ and $C$ are calibration constants depending on the sensor properties.  
 
\begin{equation}
Force = Ae^{-CV_{sensor}} + B
\label{eq_four}
\end{equation}
 
The values for the coefficients for each calibration are represented in Table~\ref{tbl:calibration-table}. 

\begin{table*}[t]\footnotesize
\centering
\makebox[\textwidth]{
\begin{tabular}{ l | lll  } 
\hline
 \textit{Sensor Number} & \multicolumn{1}{|c}{\textit{A}} & \multicolumn{1}{|c}{\textit{B}} & \multicolumn{1}{|c}{\textit{C}}  \\ \hline
   Sensor 1 & \(\displaystyle 93.01 \pm 10.04 \) &	\(\displaystyle  -95.57 \pm 14.17 \) & \(\displaystyle -0.6696 \pm 0.03006 \) \\ 
   Sensor 2 & \(\displaystyle 65.73 \pm 7.161 \) & \(\displaystyle -67.07 \pm 11.08 \) & \(\displaystyle 0.7423 \pm 0.03084 \)  \\
   Sensor 3 & \(\displaystyle 74.05 \pm 6.834 \) & \(\displaystyle -91.10 \pm 12.02 \) & \(\displaystyle -0.9516 \pm 0.03223 \) \\ 
   \hline
\end{tabular}}
\caption{Calibration constants for each of the three force pads used}
\label{tbl:calibration-table}
\end{table*}
 
\subsection{Measuring Force Distribution  }

\begin{figure}[!t]
  \includegraphics[width=\linewidth]{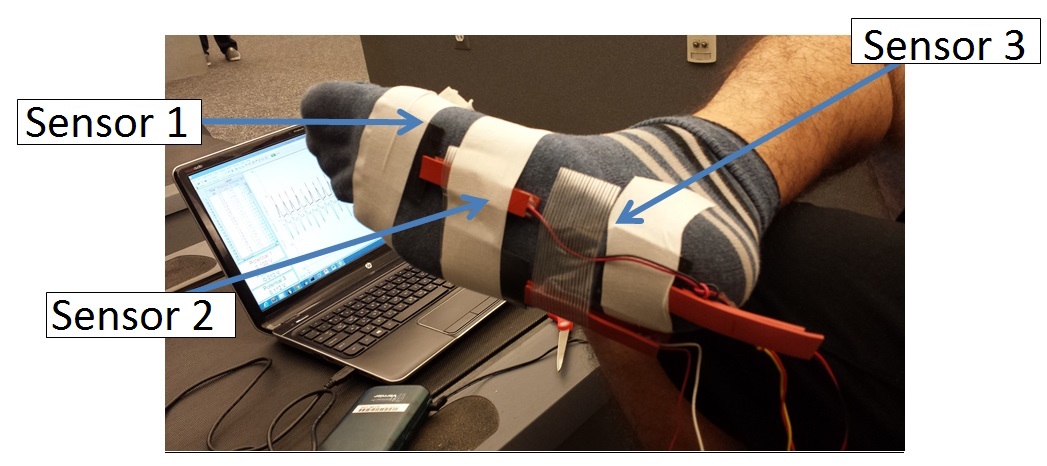}
  \caption{Sensors fixed at three different locations on the foot.}
  \label{fig:foot}
\end{figure}

The sensors were first taped onto the bottom of the foot of an RFS subject at three different locations : forefoot (Sensor \#1), midfoot (Sensor \#2) and heel (sensor \#3) . Figure~\ref{fig:foot} shows the location of different sensors  

\begin{figure}[!t]
  \includegraphics[width=\linewidth]{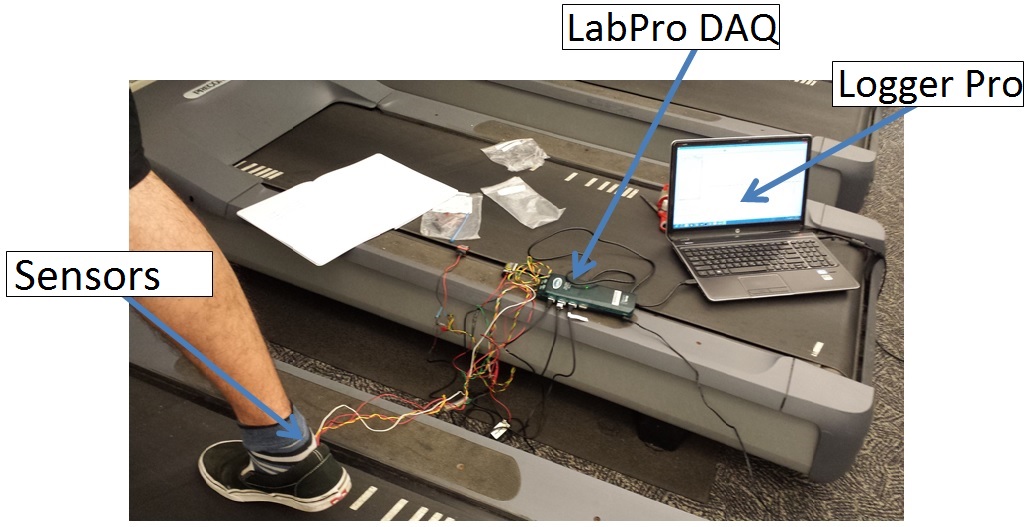}
  \caption{Experimental setup for measuring the force experienced by the foot.}
  \label{fig:setup}
\end{figure}

The sensors were next connected to the Vernier Lab Pro data acquisition interface and the voltage data was acquired in Logger Pro. Figure~\ref{fig:setup} shoes the experimental apparatus for airing force data.  

Using the calibration equations, the voltage was then converted to force, from which the force corresponding to the initial voltage measured by the sensor was subtracted.  The total force on the foot was then computed by summing the forces obtained from each sensor. 

\section{Results and Discussions}

\begin{figure}[!t]
  \includegraphics[width=\linewidth]{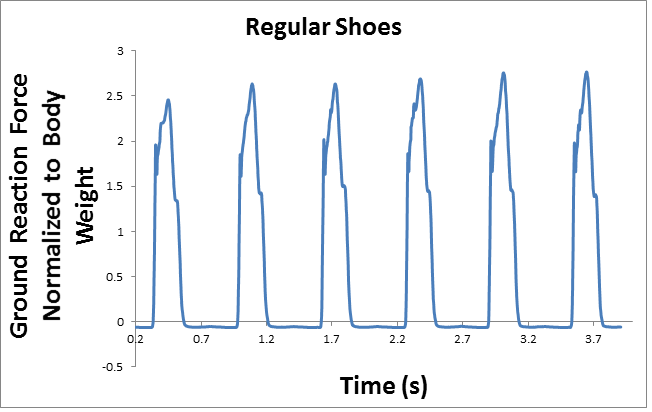}
  \caption{Graph of the VGRF normalized to the subject’s body weight during 6 cycles for a 	subject running without running shoes.}
  \label{fig:one}
\end{figure}

\begin{figure}[!t]
  \includegraphics[width=\linewidth]{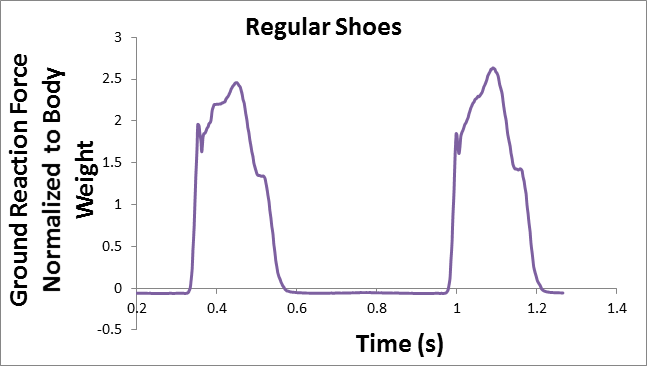}
  \caption{Graph of the VGRF normalized to the subject’s body weight during two cycles for a subject running without running shoes.}
  \label{fig:one-2}
\end{figure}

The total force was plotted as a function of time. Figures~\ref{fig:one} and ~\ref{fig:one-2} show the graphs of the ground reaction force normalized to the body weight of an RFS subject running at 9 miles per hour without running shoes. 

Assuming that maximum height of the foot is achieved at the midpoint of the swing phase, we can obtain a lower bound on $V$ by estimating the height achieved to be 20 centimeters and dividing it by half the length \( \Delta_t \) of the swing phase, thus 
 
\[ V_{foot}=\frac{2H}{\Delta_t}=1.0 m/s \pm 1.0 m/s \]
 
The previous assumption neglects the additional foot velocity due to the force generated by leg muscles. Based on the previous graph, \( M_{eff} \) was calculated during one stance period using Equation~\ref{eq_three} with a transient impulse $I$ of 0.1014 N.s , a transient impulse duration $T$ of 0.024 seconds and foot vertical velocity \( V_{foot} \) of 1 meter per second.  
 
\[ M_{eff}= (14.9 \pm 4.8)\%M_{body} \]
 
\begin{figure}[!t]
  \includegraphics[width=\linewidth]{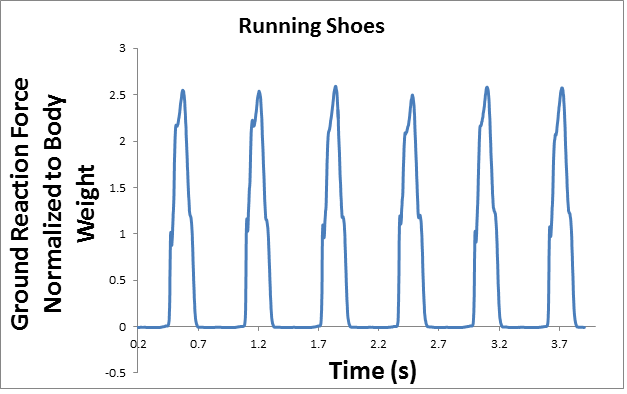}
  \caption{Graph of the VGRF normalized to the subject’s body weight during 6 cycles for a subject running with running shoes.}
  \label{fig:two}
\end{figure}

\begin{figure}[!t]
  \includegraphics[width=\linewidth]{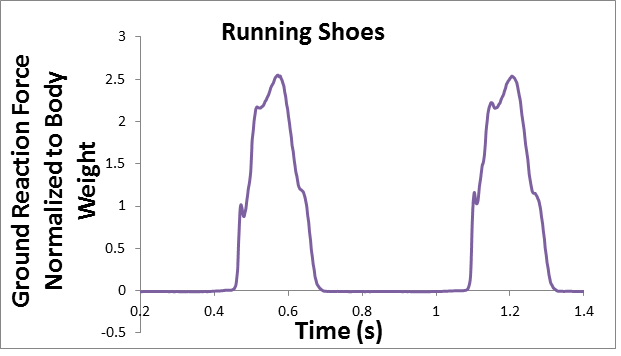}
  \caption{Graph of the VGRF normalized to the subject’s body weight during 2 cycles for a subject 	running with running shoes.}
  \label{fig:two-2}
\end{figure}

The ground reaction force was also calculated in the case of a subject wearing running shoes. Figure~\ref{fig:two} and ~\ref{fig:two-2} show the corresponding graphs. 

Using the same approach as in the case without running shoes, with a transient impulse $I$ of 0.093 N.s, a vertical foot velocity  of 1 meter per second and a transient impact time $T$ of 0.02 seconds: 

\[ M_{eff} = (7.8 \pm 1.5 )  \% M_{body} \] 

\begin{figure}[!t]
  \includegraphics[width=\linewidth]{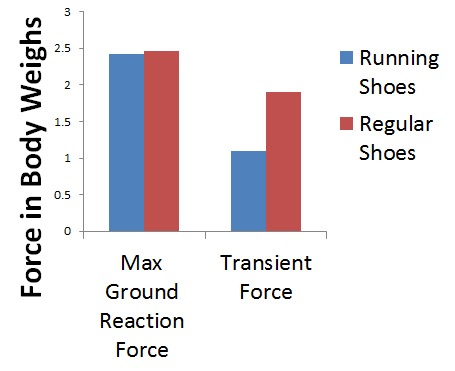}
  \caption{Histogram showing the effect of running shoes on Max Ground Reaction Force and Transient Force.}
  \label{fig:force-chart}
\end{figure}

Normalized to the body mass, the historical average for \( M_{eff} \) is 5.9\% for barefoot RFS runners \cite{lieberman2010foot}. The value 
calculated in this report for a subject wearing running shoes is in accordance with historical values. For a subject not wearing running shoes, the value is much higher because the subject is not a professional barefoot runner. These results therefore support the hypothesis that running shoes reduce transient impact forces by generating collisions with a much lower  than the ones otherwise generated. The results also show that running shoes do not affect the maximum ground reaction force experienced by the runner. Figure~\ref{fig:force-chart} shows a histogram displaying the changes observed when wearing running shoes.  

\begin{figure}[!t]
  \includegraphics[width=\linewidth]{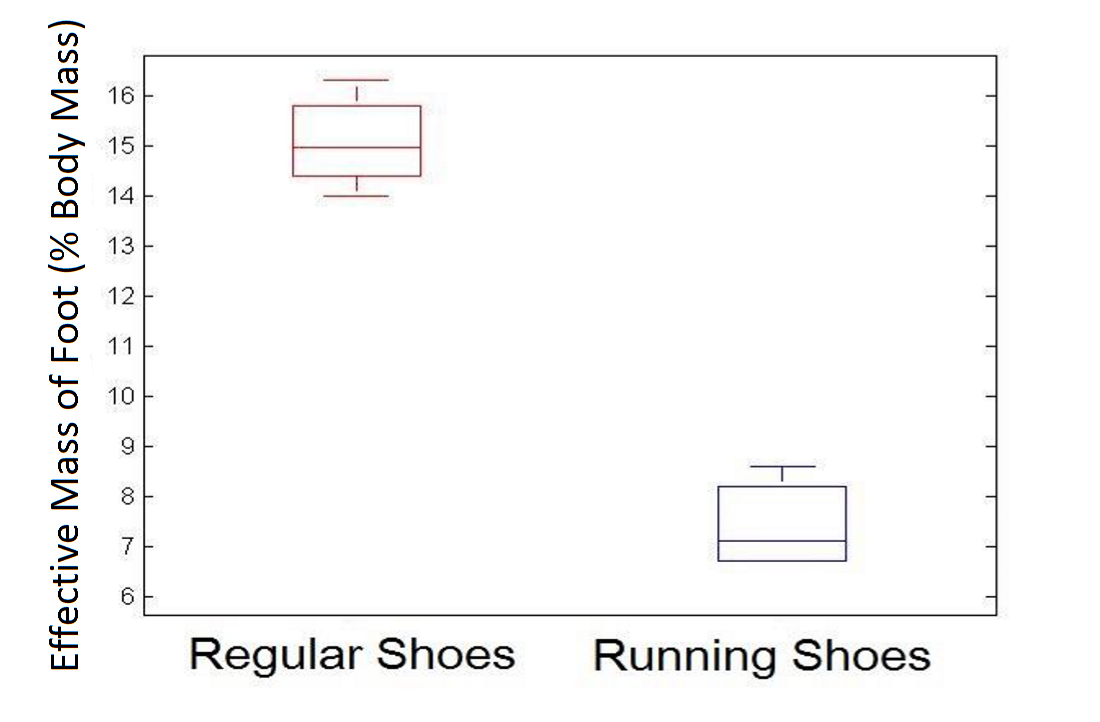}
  \caption{Boxplot showing the effect of running shoes on the foot effective mass.}
  \label{fig:boxplot}
\end{figure}

Running shoes are specifically tailored to reduce the heel impact for RFS runners by using elastic materials in a large heel to absorb part of the transient force and spread the impulse over a longer period of time \cite{lieberman2010foot}.The reduction in the effective mass is shown in Figure~\ref{fig:boxplot}.  
 
The uncertainty on  was calculated by using propagation of uncertainty. The uncertainties on the calibration coefficients were propagated into the forces measured by each sensor. The uncertainty of the force measured by each sensor was then propagated into the total force measured. Finally, the uncertainty in the total force, along with the uncertainties in the vertical velocity and impulse time, were propagated into the calculated value of \( M_{eff} \).  

\section{Conclusion}

The values obtained confirm the hypothesis that running shoes reduce the transient impact on the foot. The effective mass while running at 9 miles per hour without running shoes is \( M_{eff} = (14.9 \pm 4.8)\%M_{body} \) whereas the one with running shoes is \( M_{eff} = (7.8 \pm 1.5 )\%M_{body} \) . The main source of uncertainty comes from the force pads calibration and estimation of the foot vertical velocity just before impact.  Although the approach adopted proves that running shoes do reduce the risk of injuries and pain, it raises the question of how runners struck the ground before the invention of modern running shoes. In order to answer this question, one would analyze the foot kinematics and impact transients in long-term habitually barefoot runners and compare them to shod runners.

\bibliographystyle{asmems4}


\bibliography{asme2e}

\end{document}